# The Anisotropy of Thermal Activation Energy of 2H-NbS$_2$

Yang Wang,[1,†] Pengjian Tan,[1,†] Mingxi Chi,[1] Boxi Wei,[1] Anchun Ji,[1] Jian Wang[2,3,4,*] and He Wang[1,*]

**AFFILIATIONS**

[1]Department of Physics, Capital Normal University, Beijing 100048, China

[2]International Center for Quantum Materials, School of Physics, Peking University, Beijing 100871, China

[3]Hefei National Laboratory, Hefei 230088, China

[4]Collaborative Innovation Center of Quantum Matter, Beijing 100871, China

[†]These authors contributed equally: Yang Wang, Pengjian Tan

[*]Author to whom correspondence should be addressed: jianwangphysics@pku.edu.cn (Jian Wang); wanghe@cnu.edu.cn (He Wang)

# ABSTRACT

We investigate the anisotropic flux dynamics in 2H-NbS$_2$ single crystals through temperature-dependent resistance measurements under in-plane (H∥ab plane) and out-of-plane (**H**⊥ab plane) magnetic fields. Analysis of thermally activated flux flow (TAFF) resistance in the superconducting mixed state reveals stark contrasts in thermal activation energy ($U_0$) scaling and field dependence. For **H**⊥ab, $U_0$(0.1 T) = (1228.76 ± 53.64) K follows the Arrhenius relation with a power-law field decay ($U_0 \propto H^{-\alpha}$). In contrast, under H∥ab, $U_0$(0.5 T) = (7205.58 ± 619.65) K aligns with the modified TAFF theory, where the scaling exponent $q = 2$ potentially reflects two-dimensional vortex behavior in the TAFF region, and the field dependence of $U_0$ follows parabolic relation $(H)^\gamma[1-(H/H^*)]^2$. These results establish 2H-NbS$_2$ as a model system for probing the anisotropy of flux dynamics in layered superconductors.

# 1. INTRODUCTION

The layered transition metal dichalcogenide 2H-NbS$_2$ has emerged as a prototypical material for exploring low-dimensional superconductivity and unconventional superconducting phenomena.[1-14] Unlike other 2H-phase transition metal dichalcogenides, 2H-NbS$_2$ uniquely preserves its superconducting ground state without competing charge density wave instabilities, making it an ideal system to probe intrinsic anisotropic superconducting mechanisms.[5,6] Recent advances have uncovered rich properties of unconventional superconductivity in 2H-NbS$_2$: (i) The candidate of Ising superconductivity stabilized by strong Ising spin-orbit coupling in mono-layer, enabling critical magnetic fields far exceeding the Pauli limit in in-plane configurations, like NbSe$_2$ and MoS$_2$;[15-17] (ii) multiband superconductivity with coexisting two-dimensional (2D)-like(K-point) and three-dimensional (3D)-like (Γ-point) superconducting subbands, driven by orbital-selective pairing mechanisms,[6,11,18] and two-gap superconductivity revealed by scanning tunneling spectroscopy, with distinct coherence peaks around 0.53 meV and 0.97 meV;[2] (iii) the evidence of Fulde-Ferrell-Larkin-Ovchinnikov (FFLO) states for bulk NbS$_2$ and orbital FFLO state for flake NbS$_2$;[10,19] and (iv) the theoretic prediction of the nodal topological superconductivity under high in-plane fields, featuring Majorana flat bands and spin-triplet Cooper pairs.[20] Despite remarkable progress in understanding the unconventional superconductivity and electronic properties of 2H-NbS$_2$, studies on its magnetic flux dynamics—a critical determinant of dissipation mechanisms and high-field applicability—remain largely unexplored.

As a layered superconductor, the charge dynamics in 2H-NbS$_2$ single crystals are expected to be mostly confined to Nb-S planes, just like high-temperature cuprate superconductors.[21] This confinement due to the layered architecture can impose profound anisotropy that governs vortex lattice configurations and pinning behavior in the mixed state.[22] To probe the anisotropy of vortex flux dynamics, the transport measurements of the temperature-dependent resistance under varying field orientations (**H**⊥ab plane vs. H∥ab plane) are essential. Within the thermally activated flux flow (TAFF) region, the resistance drops enable extraction of the thermal activation energy ($U_0$),[21, 23-42] which quantifies the energy barrier against vortex motion. Crucially, the field- and orientation-dependent evolution of $U_0$ reveal two key insights: (1) The distinct temperature scaling laws for TAFF resistance—Arrhenius relation or modified TAFF method—along different orientations directly reflect anisotropic flux dynamics and potentially reveal the effective dimensionality of flux-line lattice in the TAFF region; (2) The divergent field dependencies of $U_0$ encode critical information about the dominant pinning mechanisms.

In this work, we performed systematic measurements of the temperature-dependent resistance in the TAFF region on high-quality 2H-NbS$_2$ single crystals to extract critical parameters such as the thermal activation energy $U_0$. By comparing the magnitudes of $U_0$, the temperature scaling laws required for fitting TAFF resistance, and the magnetic field dependence of $U_0$ under varying field orientations, we reveal the superconducting anisotropy of 2H-NbS$_2$ from the perspective of flux dynamics. These analyses provide crucial insights into the investigation of the anisotropic vortex-pinning mechanisms and dimensionality of flux-line lattice in the strongly anisotropic layered superconductor.

## 2. EXPERIMENTAL RESULTS

In our work, the 2H-NbS$_2$ single crystals were grown via the chemical vapor transport method, utilizing iodine as the transport agent.[43] The X-ray diffraction (XRD) data were collected using a Bruker AXS D8 Advance XRD instrument, equipped with a Cu source at a wavelength of 0.15406 nm and a power of 2.2 kW. Measurements of longitudinal resistance and Hall resistance were performed in a dry superconducting magnet system (TeslatronPT12, Oxford instruments).

2H-NbS$_2$ is a layered transition metal dichalcogenide with a hexagonal crystal structure belonging to the space group P6$_3$/mmc. Its lattice parameters are $a = b = 3.32$ Å, and $c = 11.97$ Å.[5, 44] Each unit cell contains two layers of Nb-S bonding structures, where Nb atoms form hexagonal close-packed configurations through covalent bonds with six S atoms (Fig. 1(a)). The layers are weakly bonded by van der Waals forces, which facilitate the mechanical exfoliation method to obtain 2D nanosheets, providing advantages for electronic device fabrication. In this study, room-temperature XRD was employed to characterize the 2H-NbS$_2$ single crystal. As shown in Fig. 1(b), the XRD pattern only displays (002$n$) crystallographic plane diffraction peaks, indicating high sample purity and excellent crystalline quality. The $c$-axis lattice constant was calculated to be approximately 11.96 Å, which is consistent with literature-reported values.[5, 44]

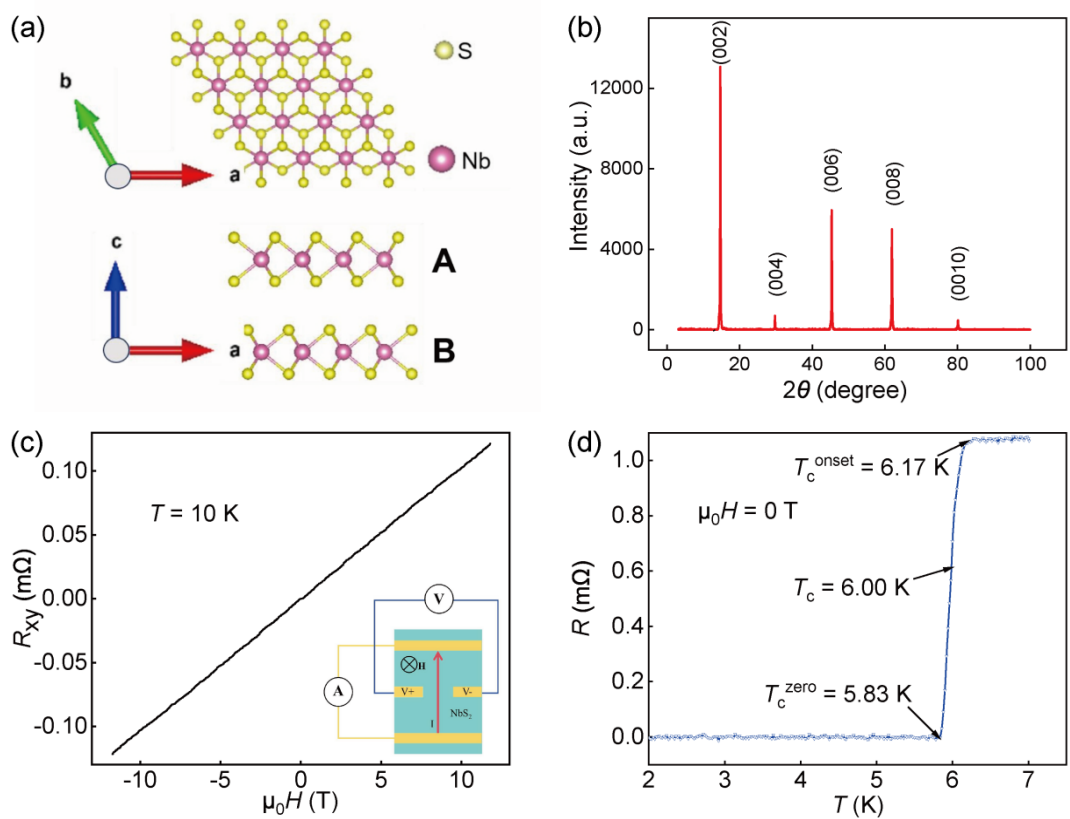

**FIG. 1.** (a) The schematic of the atomic structure of NbS$_2$. The yellow sphere represents S atoms, and the pink sphere represents Nb atoms. The upper part is the atomic structure of the ab plane, and the lower part is the atomic structure of the sideview. (b) The XRD pattern of NbS$_2$ single crystal. (c) The Hall measurements of NbS$_2$ at 10 K. Inset: schematic of the Hall and longitudinal resistance measurement configuration. (d) The temperature dependence of resistance of NbS$_2$ at μ$_0$H = 0 T. The onset transition temperature ($T_c^{onset}$) is 6.17 K, the $T_c$ is 6.00 K, and the $T_c^{zero}$ is 5.83 K.

To test the transport characteristics of 2H-NbS$_2$ single crystals, we performed Hall effect and longitudinal resistance measurements using a standard six-probe setup, as schematically shown in the inset of Fig. 1(c). For Hall measurements, we applied a DC of 8 mA and swept the out-of-plane magnetic field ($H \perp ab$) from -12 T to +12 T at a base temperature of 10 K. As demonstrated in Fig. 1(c), the Hall resistance ($R_{xy}$) exhibits a linear response with the magnetic field, yielding a positive slope that unambiguously identifies hole carriers as the dominant charge carriers in this material system. Figure 1(d) displays the temperature-dependent behavior of the longitudinal four-probe resistance $R_{xx}$ of $NbS_2$, ranging from 7 to 2 K under zero magnetic field conditions. The resistance begins to drop at 6.17 K ($T_c^{onset}$), and attains half of its normal resistance at 6.00 K (defined as $T_c$), ultimately diminishes to zero at 5.83 K ($T_c^{zero}$). These findings are consistent with the previous results, attesting to the high quality of our 2H-$NbS_2$ single crystals.[18]

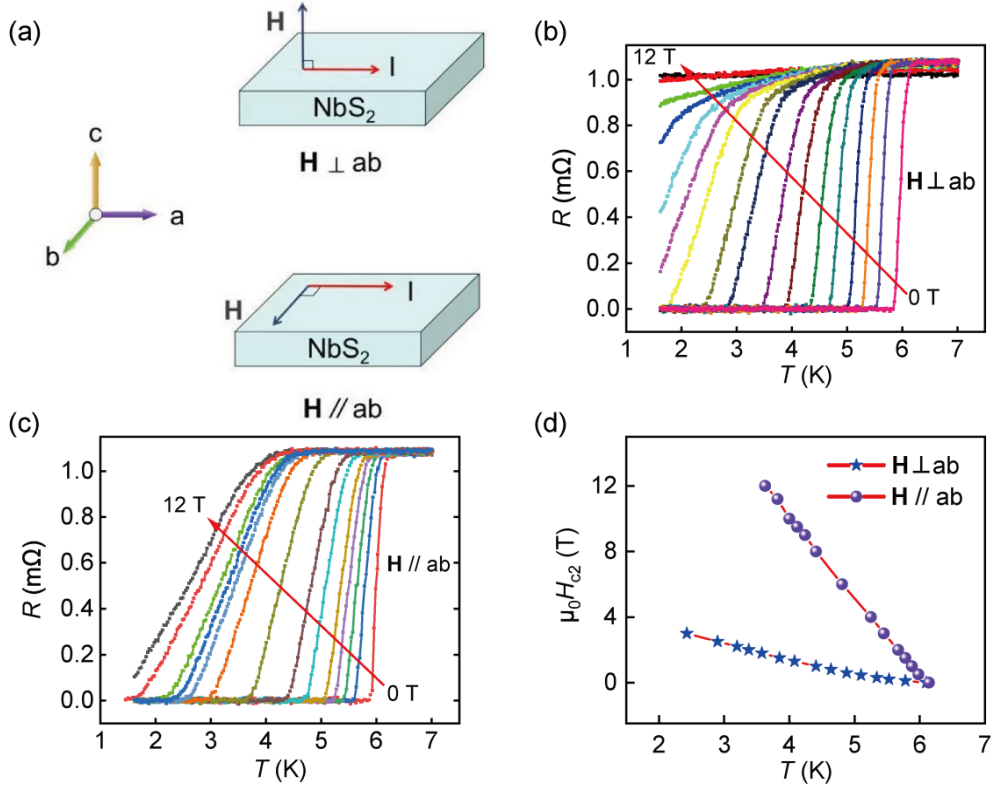

**FIG. 2.** Temperature dependence of the resistance of $NbS_2$ under the magnetic fields along different directions. (a) The upper figure: the schematic of the configuration with the magnetic field perpendicular to the *ab* plane of the sample ($H \perp ab$); The lower figure: the schematic of the setup with the magnetic field parallel to the ab plane of the sample ($H \| ab$). (b) The temperature dependence of the resistance

$R$(T) with different out-of-plane magnetic fields. The curves are measured at 0 T, 0.1 T, 0.2 T, 0.3 T, 0.45 T, 0.6 T, 0.8 T, 1.0 T, 1.3 T, 1.5 T, 1.8 T, 2 T, 2.2 T, 2.5 T, 3 T, 6 T, and 12 T, respectively. (c) The temperature dependence of the resistance $R$(T) with different in-plane magnetic fields. The curves are measured at 0 T, 0.5 T, 1 T, 1.5 T, 2 T, 3 T, 4 T, 6 T, 8 T, 9 T, 9.5 T, 10 T, 11 T, and 12 T, respectively. (d) The temperature dependence of the upper critical field $\mu_0H_{c2}$ extracted from (b) and (c).

To study the magnetic anisotropy of 2H-NbS$_2$ single crystals, we systematically measured the temperature-dependent $R_{xx}$ under both in-plane (H∥ab) and out-of-plane (**H**⊥ab) magnetic field configurations. The related measurement configurations are schematically illustrated in the upper (**H**⊥ab) and lower (H∥ab) insets of Fig. 2(a), with the magnetic field always kept perpendicular to the current direction in both cases. Figure 2(b) displays the resistance as a function of temperature $R(T)$ for **H**⊥ab from 1.6 to 7 K. A pronounced field-dependent superconducting transition is evident: As the out-of-plane magnetic field increases, superconductivity is suppressed rapidly. Notably, at the base temperature of 1.6 K, the superconducting transition cannot reach a zero-resistance state under $\mu_0H$ = 2 T, and get full suppression for $\mu_0H$ = 12 T. In striking contrast, the analogous measurements for H∥ab shown in Fig. 2(c) indicate that superconducting properties remain robust under in-plane magnetic fields. The zero-resistance state persists up to $\mu_0H$ = 11 T, and discernible superconducting features are still observable even at $\mu_0H$ = 12 T. This demonstrates the superconductivity in 2H-NbS$_2$ exhibits notable superconducting anisotropy.

To quantitatively analyze the anisotropy of superconductivity in 2H-NbS$_2$, we extracted the upper critical field $\mu_0H_{c2}$ from the $R(T)$ curves in Figs. 2(b)-(c) and plotted them against temperature in Fig. 2(d). The $\mu_0H_{c2}$ (stars) values for **H**⊥ab are obviously smaller than the values for H∥ab (solid circles), indicating that the in-plane fields have significantly reduced efficacy in suppressing superconductivity in NbS$_2$ compared to the out-of-plane magnetic fields. This anisotropic behavior clearly demonstrates that the superconducting state in 2H-NbS$_2$ exhibits strong directional dependence on the applied magnetic field, with the out-of-plane orientation being significantly more susceptible to magnetic perturbation, consistent with the previous reports on 2H-NbS$_2$ single crystal.[18]

## 3. DISCUSSION

The investigations of the TAFF resistance in superconductors establish a systematic pathway to probe vortex dynamics, superconducting mechanisms, and intrinsic material properties.[21, 30] Generally, the TAFF resistivity can be expressed as

$$\rho = (2v_0 LB/J)\exp(-J_{c0}BVL/T)\sinh(JBVL/T),$$

where $v_0$, $L$, $J$, $J_{c0}$ and $V$ denote the flux bundle hopping attempt frequency, the hopping distance, the applied current density, the critical current density, and the bundle volume, respectively. Under low-current conditions and $JBVL/T \ll 1$, the TAFF resistance can reduce to the thermally activated form:

$$R = (2R_c U/T)\exp(-U/T) = R_{0f}\exp(-U/T),$$

where $U = J_{c0}BVL$ is the thermal activation energy.[30]

Until now, there are two analytical approaches that dominate TAFF region characterization: First, the Arrhenius plot method is applicable, when $U(T, H)$ exhibits linear temperature dependence $U(H, T) = U_0(H)(1 - T/T_c)$ and the prefactor $R_{0f}$ remains constant. This leads to the logarithmic relation: $\ln R(H, T) = \ln R_0(H) - U_0(H)/T_c$, where $R_0(H)$ is determined by intercepts and $U_0(H)$ by the slope of $\ln R$ vs. $1/T$. Validity criteria for the Arrhenius plot method include i) all Arrhenius fitted lines intersect at a single point, and the temperature value corresponding to this intersection point is $T_{\text{cross}} \approx T_c^{\text{onset}}$; ii) the dependence of $\ln R_0$ on the $U_0$ keeps linear, and the reciprocal of the linear slope is approximately $T_c^{\text{onset}}$. Second, the modified TAFF method can be employed for nonlinear temperature dependencies expressed as $U(H, T) = U_0(H)(1 - T/T_c)^q$. The corresponding resistivity relation becomes

$$\ln R = \ln(2R_c U_0) + q\ln(1 - T/T_c) - \ln T - U_0(1 - T/T_c)^q/T,$$

where $R_c$ and $U_0$ are temperature-independent parameters.[30] For $q = 1$, the modified

TAFF relation is reduced to the Arrhenius relation. Generally, $q = 1.5$ and $q = 2$ were observed in high-temperature superconductors showing 3D and 2D behavior, respectively.[33] Hence, both the Arrhenius plot and the modified TAFF method provide useful tools for advancing our understanding of vortex dynamics in the superconducting mixed state of $NbS_2$.

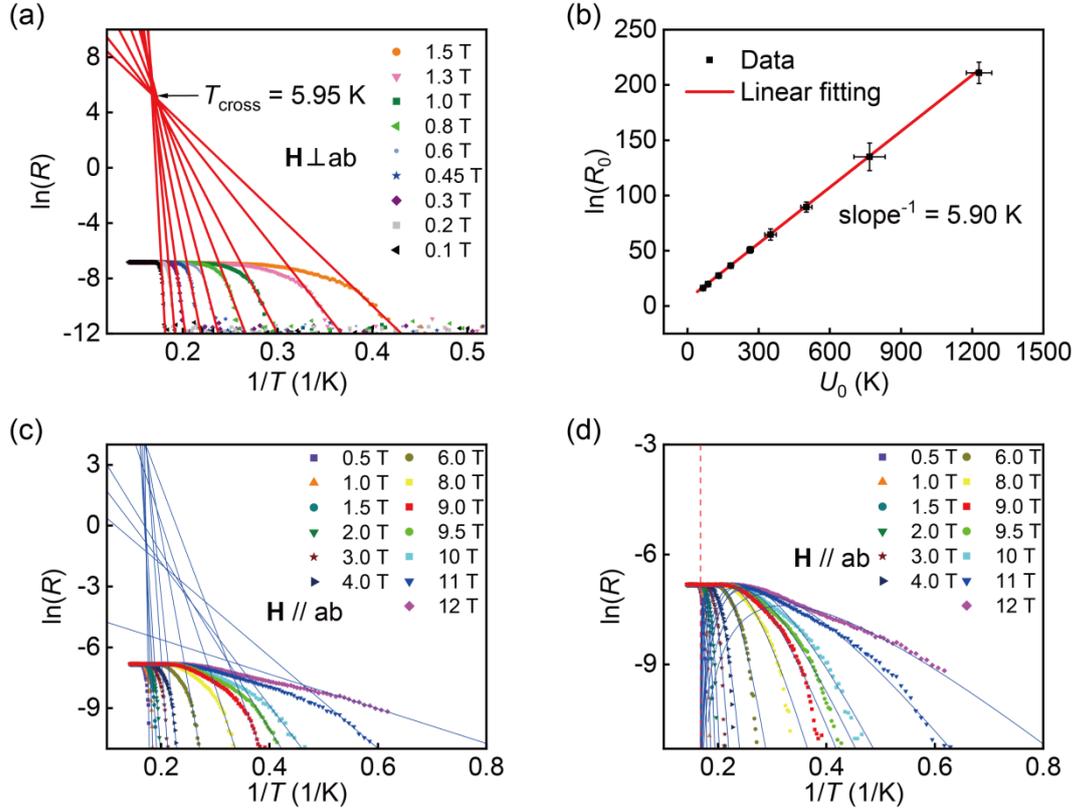

**FIG. 3.** (a) Arrhenius plot of $R(T)$ curves from Fig. 2(b) under the out-of-plane magnetic fields $\mu_0 H = $ 0.1, 0.2, 0.3, 0.45, 0.6, 0.8, 1.0, 1.3 and 1.5 T, respectively. The solid lines are linear fit to the TAFF region. (b) The linear fit to $U_0$ versus $\ln(R_0)$ curve derived from Fig. 3(a). (c) Arrhenius plot of $R(T)$ curves from Fig. 2(c) under the in-plane magnetic fields $\mu_0 H = $ 0.5, 1, 1.5, 2, 3, 4, 6, 8, 9, 9.5, 10, 11 and 12 T, respectively. The solid lines are linear fit to the TAFF region. (d) The fitting results of $R(T)$ curves from Fig. 2(c) by using the modified TAFF method. The blue solid lines are the fitting curves of the modified TAFF formula.

Figure 3 further reveals distinct behaviors of $NbS_2$ single crystals under out-of-plane versus in-plane magnetic fields within the TAFF region. For the out-of-plane magnetic field, the solid lines for Arrhenius plots linearly fit the experimental data well with a single intersection point $T_{cross}$ = 5.95 K, very close to the superconducting onset transition temperature $T_c^{onset}$ = 6.17 K, satisfying the Arrhenius criterion. The activation energy $U_0$ is extracted from the slopes of these lines, and Fig. 3(b) shows a linear relationship between $\ln R_0$ and $U_0$ with a slope of 5.90 K—again near the $T_c^{onset}$, confirming the Arrhenius method's validity for out-of-plane fields. However, under the in-plane fields, as shown in Fig. 3(c), the blue solid lines fail to converge at a single point, violating the Arrhenius relation and highlighting its limited applicability. To overcome this, a modified TAFF analysis (blue curves in Fig. 3(d)) is employed, achieving excellent agreement with experimental data via the parameter fitting. Notably, the key parameter $q$ for $NbS_2$ single crystal is 2, which is usually identified as a hint of 2D vortex behavior in high-temperature superconductors.[30-33, 38]

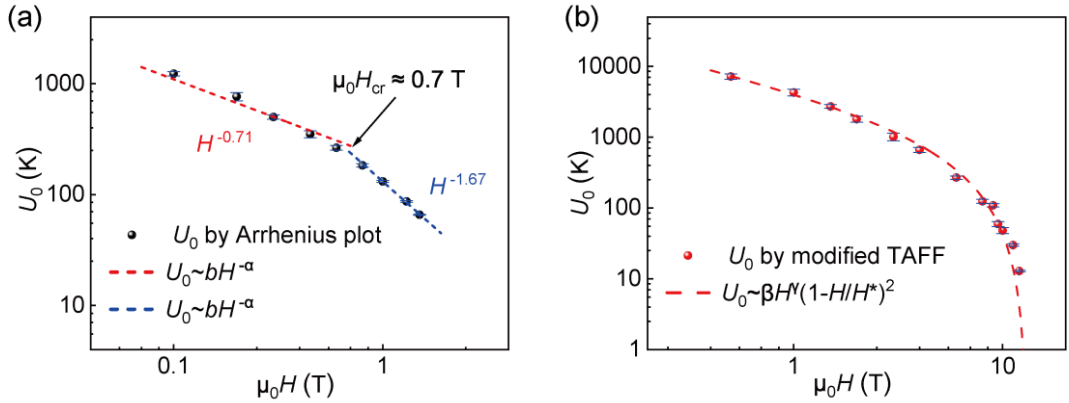

**FIG. 4.** Thermally activation energy is plotted against magnetic field $\mu_0 H$ along the out-of-plane direction (a) and in-plane direction (b) on a log-log scale. The dashed curves (lines) are the fitting results.

Figure 4 presents the magnetic field dependence of the thermal activation energy in $NbS_2$. As demonstrated in Fig. 4(a), $U_0$ extracted from Arrhenius plots at various out-of-plane magnetic fields exhibit significant field-dependent variations. For $\mu_0 H$ = 0.1 T, the $U_0$ is about (1228.76 ± 53.64) K. With increasing the magnetic field, a power-law relationship between $U_0$ and $H$ is clearly expressed by $U_0(H) \sim H^{-\alpha}$, where $\alpha$ is the fitting parameter. The $U_0$ exhibits a dependence of $H^{-0.71}$ in the regime of relatively low magnetic fields, whereas it shows a dependence of $H^{-1.67}$ in the regime of relatively high magnetic fields. The two regimes are clearly visible with a crossover occurring at a characteristic field $H_{cr} \approx 0.7$ T. Similar crossover between two power-law dependence of $U_0(H)$ has been observed in several systems[27, 30, 32-35, 37, 38, 45], including high-temperature cuprate superconductors, ion-based superconductors, and transition metal transition metal dichalcogenides. The vortex pinning mechanism can be preliminarily deduced from the magnitude of the exponent α. In the low-field regime ($\mu_0H < 0.7$ T), the observed $α = 0.71$ deviates significantly from the theoretical predictions for single-vortex pinning ($α \approx 0$) and plastic weak pinning ($α = 0.5$),[21, 25, 35] suggesting the presence of plastic strong pinning. Notably, prior work proposes that the collective pinning for the vortex may coexist with plastic pinning in this low-field regime.[37] In contrast, under high magnetic fields ($α = 1.67$), the accelerated decrease in activation energy $U_0$ with increasing magnetic field is linked to the formation of entangled vortex liquid states within regions governed by point-defect-dominated plastic strong pinning.

The magnetic field dependence of the activation energy $U_0$ for $NbS_2$ under in-plane fields is summarized in Fig. 4(b). At $\mu_0H = 0.5$ T, $U_0$ reaches approximately $(7205.58 \pm 619.65)$ K, significantly exceeding the $U_0$ at $\mu_0H = 0.1$ T [$\approx (1228.76 \pm 53.64)$ K] observed under out-of-plane fields (Fig. 4(a)). Unlike the power-law dependence of $U_0(H)$ for out-of-plane fields, the in-plane-field data are well described by a parabolic function: $U(H) = \beta(\mu_0H)^\gamma[1-(H/H^*)]^\delta$, where $H^*$ denotes the Kramer's scaling field.[46] The term $\beta(\mu_0H)^\gamma$ reflects a reduction in critical current density $J_c$, consistent with collective flux creep dynamics, while $[1-(H/H^*)]^\delta$ signifies the progressive suppression of superconducting regions with increasing $H$. For $\beta = 4653.88 \pm 81.68$ K, $\gamma = -0.77 \pm 0.028$, and $\delta = 2$, the fitting yields an excellent agreement with the goodness of fit $R^2=0.996$, capturing the parabolic $U_0(H)$ behavior. The origin of this parabolic dependence remains unclear. Notably, similar $U_0(H)$ trends have been reported in other systems.[28, 29, 33, 38] For example, in polycrystalline $MgB_2$ film, the parabolic behavior is ascribed to plastic deformation of vortex lattice induced by pre-existing quenched dislocations.[28] Conversely, in $NbSe_2$ and $Fe_{0.0015}NbSe_2$, it indicates the elastic deformation of flux lines.[38]

## 4. CONCLUSION

We systematically investigated the temperature-dependent resistance of 2H-NbS$_2$ under the in-plane and out-of-plane magnetic fields, respectively. This comparative analysis reveals a pronounced anisotropy of $\mu_0 H_{c2}$ with the value for H∥ab much larger than that for **H** ⊥ ab at $T \ll T_c$. Analysis of the TAFF region further demonstrates pronounced anisotropy in the following aspects: First, the thermal activation energy under out-of-plane fields, $U_0(0.1\ \text{T}) = (1228.76 \pm 53.64)$ K, is significantly smaller than that under in-plane fields, $U_0(0.5\ \text{T}) = (7205.58 \pm 619.65)$ K. Second, distinct temperature scaling laws govern the TAFF region—the condition under out-of-plane fields follows an Arrhenius relation (corresponding to $q = 1$), while that under in-plane fields requires a modified TAFF model with $q = 2$ for optimal fitting, potentially revealing the 2D-characteristics of vortices in the TAFF region. The behavior that the fitting parameter $q$ changes from 1 for **H**⊥ab to 2 for H∥ab is analogous to observations in the Fe$_{1+y}$(Te$_{1+x}$S$_x$)$_z$ sample.[32] Third, the thermal activation energy exhibits marked directional dependence on magnetic fields: under out-of-plane fields, the activation energy shows power-law dependence on the magnetic field, $U_0 \sim H^{-\alpha}$. The exponent $\alpha$ larger than 0.5 under low magnetic fields indicates the presence of the plastic strong pinning. Conversely, the thermal activation energy follows a parabolic relationship with the in-plane magnetic field, $U(H) = \beta(\mu_0 H)^{\gamma}[1-(H/H^*)]^2$, the detailed mechanism has not been well understood so far. These findings provide new experimental insights into the quasi-2D flux dynamics of 2H-NbS$_2$, emphasizing the critical role of crystallographic orientation in modulating both magnetic field responses and competing pinning mechanisms.

## DATA AVAILABILITY STATEMENT

The data that support the findings of this study are available from the corresponding author upon reasonable request.

## ACKNOWLEDGMENTS

This work was financially supported by the National Natural Science Foundation of China [Grant No. 12488201 (J.W.); Grant No. 62175169 (A.J.)], the Innovation Program for Quantum Science and Technology [2021ZD0302403 (J.W.)], Beijing Natural Science Foundation [Grant No. 1202005 (H.W.)], Capacity Building for Sci-Tech Innovation-Fundamental Scientific Research Funds [Grant No. 20530290057 (H.W.)], the Science and Technology Project of Beijing Municipal Education Commission [Grant No. KM202010028014 (H.W.)].